\def\asmz{\alpha_S(M_Z^2)}

\documentclass{ws-procs10x7a}

\begin{document}

\title{QCD Theory}

\author{W. J. Stirling}

\address{Institute for Particle Physics Phenomenology, University of Durham, \\
Durham DH1~3LE, UK\\E-mail: w.j.stirling@durham.ac.uk}

\twocolumn[\maketitle\abstract{Quantum Chromodynamics is an established part of the Standard Model and
an essential part of the toolkit for searching for new physics at high-energy colliders. I present a status report on the
theory of QCD and review some of the important developments in the past year.}]

\section{Introduction}\label{sec:intro}

Quantum Chromodynamics (QCD), the gauge field theory that describes the 
interactions of coloured quarks and gluons, is one of the components
of the SU(3)$\times$SU(2)$\times$U(1) Standard Model. At short distances,
equivalently high energies, the effective coupling is small and the theory
can be studied using perturbative techniques. Enormous effort has been made over the past three decades to calculate higher-order pQCD corrections to
phenomenologically relevant quantities and, as we shall see in this review,
 the frontier is now at next-to-next-to-leading order (NNLO). Non-perturbative
QCD contributions are also important. They control, for example, the transitions between hadrons
and quarks and gluons, both in the initial and final states, and thus have a key role to play in
dictating the overall `shape' of a high-energy collider event. Again, there has been  much progress
in understanding and modelling these non-perturbative effects in recent years, including approaches based
on lattice calculations, Regge theory, Skyrme and large-$N_c$ models etc., to the extent that
in many areas of application QCD phenomenology is now a high-precision science. However there are still gaps
in our understanding, in particular there are processes for which our knowledge is still only at the semi-quantitative level and much
more work needs to be done. Examples are semi-hard, exclusive and soft processes at colliders. This is another important frontier for QCD.

`QCD as a high-precision tool at colliders' is the theme of this talk. I will begin by reviewing the status
of pQCD calculations and $\alpha_S$ measurements, and then highlighting some significant calculational developments in the past year. This includes
the emergence of a new way of doing pQCD scattering amplitude calculations that could have a dramatic impact on
QCD phenomenology. I will also discuss some important and novel LHC processes where more calculational effort is needed.
Lack of space prevents a detailed review of all the interesting experimental results on QCD-related processes presented
at this Conference, but many of these are covered in other plenary talks.

\section{Status of pQCD calculations}\label{sec:status}

For a broad class of `hard' high-energy processes involving hadrons (i.e. suitably inclusive, with at least one large momentum transfer scale), the cross section can be calculated in QCD perturbation theory:\cite{QCDBOOK}
\begin{equation}
d\sigma = A \alpha_S^N [\;  1 + C_1 \alpha_S + C_2 \alpha_S^2 + ...\; ]
\label{eq:one}
\end{equation}
The classic example is the three-jet cross section in $e^+e^- \to\; $hadrons, for which $N=1$.  The $A$, $C_1$ and $C_2$ coefficients define the LO, NLO and NNLO perturbative contributions respectively. Where the hard process involves two large but unequal energy scales, the perturbative coefficients may be dominated by large logarithms of the two scales,
\begin{eqnarray}
d\sigma &=& A \alpha_S^N [\; 1 + (c_{1,1}L + c_{1,0}) \alpha_S  \nonumber \\
&+& (c_{2,2}L^2 + c_{2,1} L +  + c_{2,0}) \alpha_S^2 + ...\; ]
\label{eq:two}
\end{eqnarray}
where $L = \log(M/q_T),\; \log(1/x), \; \log(1-T), \; ... \gg 1$, for example. In this case one can often resum the leading and some subleading logarithms to all orders to improve the perturbative prediction, the  $c_{n,n}$, $c_{n,n-1}$, $c_{n,n-2}$ coefficients
defining the  LL, NLL, NNLL contributions respectively. We will show examples of state-of-the-art resummed pQCD phenomenology below. 

\begin{figure}%1
\epsfxsize200pt
\figurebox{}{}{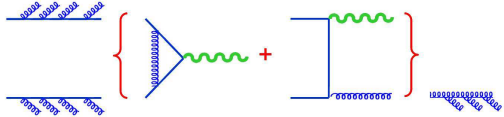}
%\vspace*{-0.5cm}
\caption{Interfacing a NLO pQCD calculation with parton showering in the production of a $W$ or $Z$ boson in hadron-hadron collisions.}
\label{fig:NLOPS}
\end{figure}
Perturbative QCD calculations become more complex the more external quarks and gluons are involved. Thanks to the development of automated codes such as MADGRAPH\cite{MADGRAPH} and CompHEP\cite{COMPHEP}, in principle {\it any} tree-level (leading-order) scattering amplitude can be calculated.
However the large renormalization scale dependence generated by the overall $\alpha_S(\mu^2)^N$ factor means that precision
predictions for overall rates are not possible at this order. This is partially solved by extending the calculation to NLO, where the coefficient $C_1$ receives real and virtual contributions: $d\sigma_V^{(N)} + d\sigma_R^{(N+1)}$. Although scale dependence is reduced  (for a cross section calculated in pQCD at order $\alpha_S^N$, $\mu^2 {\rm d}\sigma_N /{\rm d} \mu^2
= O(\alpha_S^{N+1})$)
it still dominates the uncertainty on many NLO $\alpha_S$ measurements. Another improvement at NLO
is that jet structure
begins to emerge, i.e. a jet can contain one or two partons and therefore has a perturbative `width', in contrast to LO where jet = parton. Indeed an important advance in NLO technology has been the development of techniques for interfacing the perturbative NLO calculation with parton shower models (i.e. leading logarithm parton branching followed by hadronization), illustrated schematically in Fig.~\ref{fig:NLOPS}. The procedure for doing this is highly non-trivial --- one must avoid
double counting gluon emissions, and deal with cancelling negative and positive singularities from the
two types of fixed-order contribution shown in $\{\,\}$ in Fig.~\ref{fig:NLOPS}.

\begin{figure}%1
\epsfxsize180pt
\figurebox{}{}{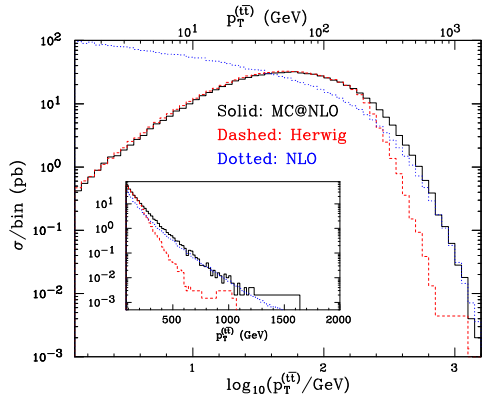}
%\figurebox{}{}{nnllpaper.ps}
\caption{MC@NLO\protect\cite{Frixione:2003ei} prediction for the $t\bar t$ transverse momentum distribution at the Tevatron collider.}
\label{fig:top_lhc_ptpair}
\end{figure}

The most advanced formalism of this type is the MC@NLO programme of Frixione {\it et al.}\cite{Frixione:2003ei}, which combines
the full pQCD NLO calculation with the HERWIG parton shower Monte Carlo. MC@NLO already contains a number of important processes for hadron collider phenomenology, including  $W$, $Z$, $WW$, $WZ$, $ZZ$, $b\bar b$, $t\bar t$
and  $H$ production. The benefits
of combing fixed order N$^n$LO and parton showers are obvious: one retains the correct overall rate (with reduced scale dependence) and the full hard-scattering kinematics,  while generating a complete event picture and a consistent treatment of collinear logarithms to all orders. This is illustrated in Fig.~\ref{fig:top_lhc_ptpair}, which shows the MC@NLO prediction for the $p_T$ distribution of $t \bar t$ pairs
at the Tevatron $p \bar p$ collider. One would expect the fixed-order and the parton shower (resummed leading logarithm) approaches to give the correct 
prediction at low and high $p_T$ respectively, and indeed the MC@NLO prediction interpolates smoothly between them.

In contrast to LO, there is as yet no automated technique for NLO calculations -- 
each process has to be treated individually -- although this is currently a topic of detailed study. As a result, there
are many processes for which the full NLO corrections are not yet known. The most phenomenologically important are those involving the production of multiple gauge bosons and heavy quarks in hadron collisions, which form the background to many New Physics processes.\cite{Campbell}
For example, the $O(\alpha_S^4)$ $q\bar q, gg \to t \bar t b \bar b$ process forms an irreducible background to 
associated $t \bar t H(\to b \bar b)$ production at the LHC. The signal to  background ratio is $O(1)$, but the latter
is currently known only at LO and there is therefore a large scale dependence in the prediction.

\section{$\alpha_S$ measurements}\label{sec:alphas}

\begin{figure}%1
\epsfxsize130pt
\figurebox{}{}{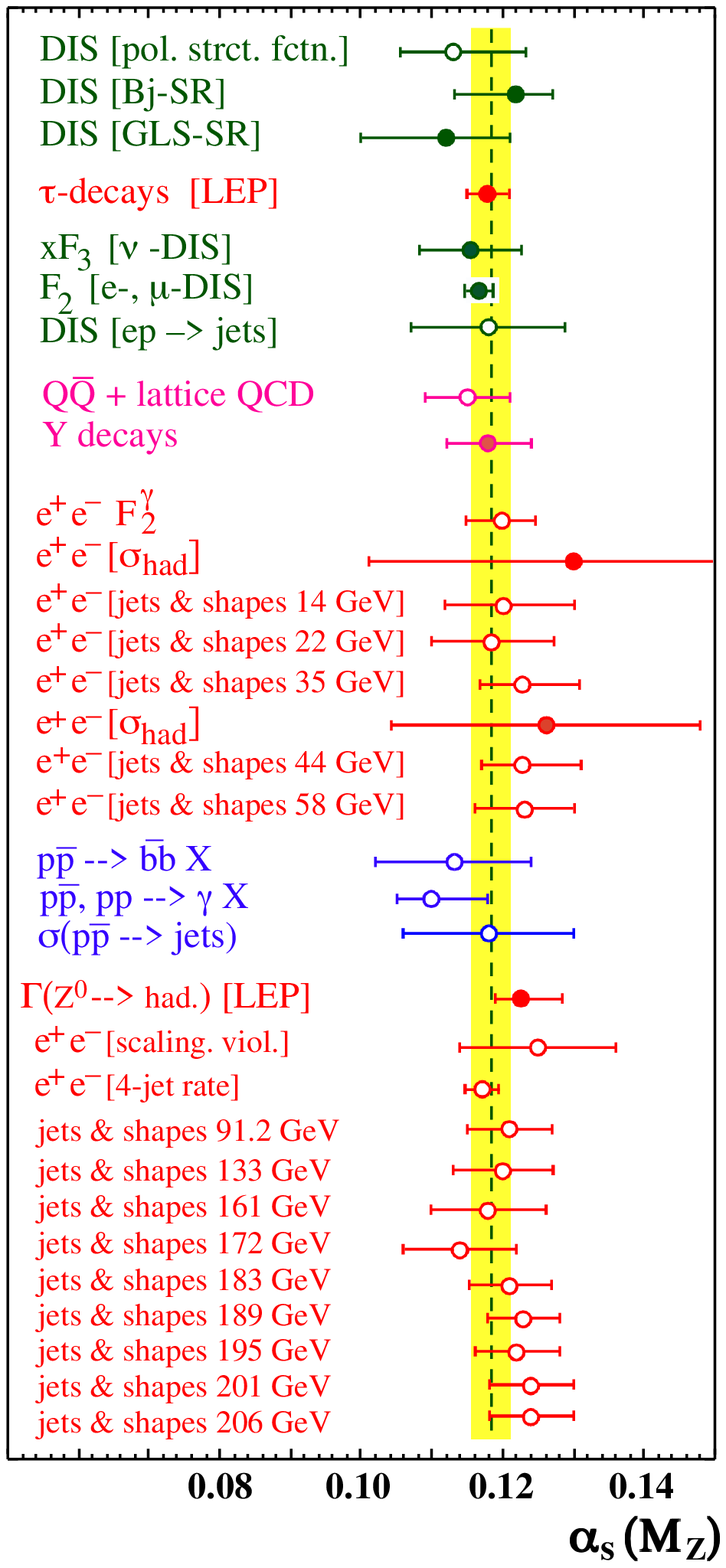}
\caption{Summary of $\asmz$ measurements, from S.~Bethke.\protect\cite{Bethke:2004uy}}
\label{fig:asmz04}
\end{figure}

Figure~\ref{fig:asmz04}, from S.~Bethke,\cite{Bethke:2004uy} summarizes the $\asmz$ measurements from some of
the most accurate recent determinations. The table contains a mix of NLO and NNLO measurements.
For experiments performed
at energy scales different from $M_Z$, the $\alpha_S$ values measured
at $\mu^2 = Q^2_{\rm exp}$ are 
converted to $\asmz$ using the standard expressions.\cite{QCDBOOK} The consistency of 
the various measurements is remarkable --- $\alpha_s$ is indeed a 
universal parameter! Defining a `world average' value presents
a technical difficulty, however.
Since the errors on most of the measurements are largely
theoretical --- often based on estimates of unknown higher-order
corrections or non--perturbative effects --- and neither Gaussian
nor completely independent, the overall error on the combined
value of $\asmz$ cannot be obtained from standard statistical
techniques, see Ref.~\cite{Bethke:2004uy} for details. Bethke's (2004) world average value and error is
\begin{equation}
\alpha_S^{\overline{\rm MS}, {\rm NNLO}}(M_Z^2) = 0.1182 \pm  0.0027
\label{eq:asmz}
\end{equation}
In view of the 
consistency of all the measurements, and in particular of those
with the smallest uncertainties, it seems unlikely that
future `world average' values of $\alpha_S$ will deviate significantly, if at all,
from the current value given in (\ref{eq:asmz}). Indeed, the corresponding world average value in 
2002 was almost identical at $0.1183 \pm 0.0027$.\cite{Bethke:2004uy}

At this Conference, a number of new $\alpha_S$ measurements have been reported, from processes 
including  (i) HERA jet cross sections and shape variables, (ii) 4-jet shape and jet shape moments
from reanalysed JADE ($e^+e^-$) data, and (iii) LEP jet shape observables and 4-jet rates. All these
are NLO measurements, and all are consistent with the values shown in Fig.~\ref{fig:asmz04}.

\section{NNLO}\label{sec:NNLO}

As experimental present and future measurements reach the few per cent accuracy level, pQCD calculations at
NNLO or higher are required. NNL0 is in fact the current perturbative frontier, although NNLO corrections for a number of processes, particularly inclusive quantities with an external electroweak gauge boson providing the large energy scale,
for example the $e^+e^-\to\ $hadrons total cross section, have been known for some time.\cite{QCDBOOK}
\begin{table*}[t]
\begin{center}
\begin{tabular}{|c|l|}  
\hline
\raisebox{0pt}[10pt][5pt]{} & DIS polarised and unpolarised structure function coefficient functions \\
\raisebox{0pt}[10pt][5pt]{$ep$} & Sum Rules (GLS, Bj, ...) \\
\raisebox{0pt}[10pt][5pt]& DGLAP splitting functions\protect\cite{Moch:2004pa,Vogt:2004mw} \\
\hline
\raisebox{0pt}[10pt][5pt]{} & total hadronic cross section, and $Z\to\; $hadrons, $\tau \to \nu +\; $hadrons \\
\raisebox{0pt}[10pt][5pt]{$e^+e^-$} & heavy quark pair production near threshold \\
\raisebox{0pt}[10pt][5pt]{} & $C_F^3$ part of $\sigma(3\; {\rm jet})$\protect\cite{Gehrmann-DeRidder:2004xe} \\
\hline
\raisebox{0pt}[10pt][5pt]{}& inclusive $W$, $Z$, $\gamma^*$\protect\cite{Hamberg:1990np,Harlander:2002wh}\\
\raisebox{0pt}[10pt][5pt]{} & inclusive $\gamma^*$ with longitudinally polarised beams\protect\cite{Ravindran:2003gi}\\
\raisebox{0pt}[10pt][5pt]{$pp$} &  $W$, $Z$, $\gamma^*$ differential rapidity distribution\protect\cite{Anastasiou:2003ds,Anastasiou:2003yy}\\
\raisebox{0pt}[10pt][5pt]{} & $H$, $A$ total\protect\cite{Harlander:2002wh,Harlander:2002vv,Anastasiou:2002yz,Anastasiou:2002wq,Ravindran:2003um}
and differential rapidity distribution\protect\cite{Anastasiou:2004xq}\\ 
\raisebox{0pt}[10pt][5pt]{}& $WH$, $ZH$\protect\cite{Brein:2003wg} \\
\hline
\raisebox{0pt}[10pt][5pt]{HQ} & $Q\overline Q$--onium and $Q\bar q$ meson decay rates\\
\hline\end{tabular}
\caption{Summary of quantities known (exactly) to NNLO in pQCD. References are given to
recent calculations only. Not included are other partial or approximate (e.g. soft, collinear) results and 
NNLL improvements. \label{tab:nnlo}}
\end{center}
\end{table*}

\begin{figure}%1
\epsfxsize200pt
\figurebox{}{}{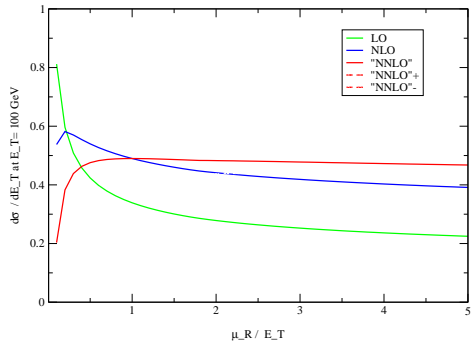}
\vspace*{-0.8cm}
\caption{Dependence of the predictions for the Tevatron jet inclusive cross section at $E_T = 100$~GeV on the 
renormalization/factorization scale $\mu$ in the $\overline{\rm MS}$ scheme, from 
Ref.~\protect\cite{Glover:2002gz}}
\label{fig:joey}
\end{figure}

Table~\ref{tab:nnlo} summarises the quantities for which the NNLO pQCD corrections are currently known. The calculations of those processes with references listed have all been completed within the past two years. Perhaps the most important NNLO calculation still outstanding is the inclusive high-$E_T$ jet distribution at hadron colliders. This is needed to complete the 
ingredients for a full NNLO parton distribution function (pdf) global analysis (see below), and to search for a New Physics
contribution to the high-$E_T$ tail of the distribution. Schematically,
\begin{eqnarray}
\frac{d\sigma^{\rm jet}}{d E_T} &=&  a A 
+ a^3 \big( B + 2b_0 LA\big) + a^4 \big(C   \nonumber \\
&& + 3b_0 LB + (3b_0^2 L^2 + 2 b_1 L)A \big) 
\label{eq:four}
\end{eqnarray}
with $ a = \alpha_S(\mu_R^2)$ for renormalization and factorization scale $\mu_R$.
Although the NNLO correction coefficient $C$ is not yet known, its likely effect in reducing factorization and renormalization scheme dependence is illustrated in Fig.~\ref{fig:joey}\cite{Glover:2002gz}, where the cross section is plotted for the 
choices $C=0,\; \pm B^2/A$. 
%For these 'reasonable' choices, the scale dependence is reduced to below the $10\%$ level over
%a broad range of scale values.
%

\begin{table*}[t]
 \begin{eqnarray} 
{\partial q_i(x,Q^2) \over \partial \log Q^2} &=&  {\alpha_S\over 2\pi}
\int_x^1 \frac{dy}{y} \; 
\Big\{
    P_{q_i q_j}(y,\alpha_S) q_j(\frac{x}{y},Q^2) 
  \;   + \; 
       P_{q_i g}(y,\alpha_S) g(\frac{x}{y},Q^2)
    \Big\} \nonumber \\
{\partial g(x,Q^2) \over \partial \log Q^2}& = & {\alpha_S\over 2\pi} \int_x^1
\frac{dy}{y}\; 
\Big\{
    P_{g q_j}(y,\alpha_S) q_j(\frac{x}{y},Q^2) \;   +  \; 
    P_{g g}(y,\alpha_S) g(\frac{x}{y},Q^2)
    \Big\} 
\label{eq:dglap}
\end{eqnarray}
\end{table*}
The complexity of the NNLO jet calculations stems from the fact that the singular parts of the different components of the $O(\alpha_S^4)$ calculation (the 2--loop, 2--parton final state; the 1--loop--squared, 2--parton final state, the 1--loop, 
3--parton final state; the tree-level, 4--parton final state) have to be identified and cancelled, leaving behind a non-singular net contribution that can be evaluated numerically to yield the coefficient $C$ in (\ref{eq:four}). The key is to identify and calculate the various `subtraction terms' that add and subtract to render the loop (analytically) and real emission (numerically) contributions separately finite, see the review by Glover\cite{Glover:2002gz} for a more detailed discussion and list of references to recent work in this area. 

\section{Three-loop splitting functions}
Hard-scattering cross sections in hadron-hadron collisions (e.g. $pp\to H+ X$) are obtained by convoluting subprocess cross sections ($\hat\sigma$) with parton distribution functions, $f_i(x,Q^2)$, whose factorization-scale ($Q^2$) dependence 
is determined by the DGLAP evolution equations (\ref{eq:dglap}). Consistency requires the subprocess cross sections and DGLAP splitting functions to be calculated to the same order in perturbation theory: 
\begin{eqnarray}
P(x,\alpha_S) &=& P^{(0)}  +  \alpha_S P^{(1)}  +  \alpha_S^2 P^{(2)}  + ... \nonumber \\
\hat\sigma &=& \hat\sigma^{(0)}  +  \alpha_S \hat\sigma^{(1)}  +  \alpha_S^2 \hat\sigma^{(2)}  + ...
\label{eq:six}
\end{eqnarray}
The full set of LO and NLO splitting functions were calculated in the 1970s and 1980s -- see Ref.~\cite{QCDBOOK} for
details. Calculations of the NNLO coefficient functions of the various hadron collider subprocess cross sections date back to the early 1990s, see Table~\ref{tab:nnlo}. However it was only a few months ago that the full set of NNLO (`three-loop') splitting functions finally became available when, in a landmark calculation, 
Moch, Vermaseren and Vogt\cite{Moch:2004pa,Vogt:2004mw} (MVV) completed the determination of all the components of the 
full splitting function matrix at NNLO, i.e. the $P^{(2)}_{q_iq_j}$, $P^{(2)}_{q_i g}$, $P^{(2)}_{g q_i}$ and $P^{(2)}_{gg}$ functions of (\ref{eq:dglap}), (\ref{eq:six}). Both (very lengthy) complete analytic forms and (compact) numerical approximations for practical applications are provided.
\begin{figure}%1
\epsfxsize160pt
\figurebox{}{}{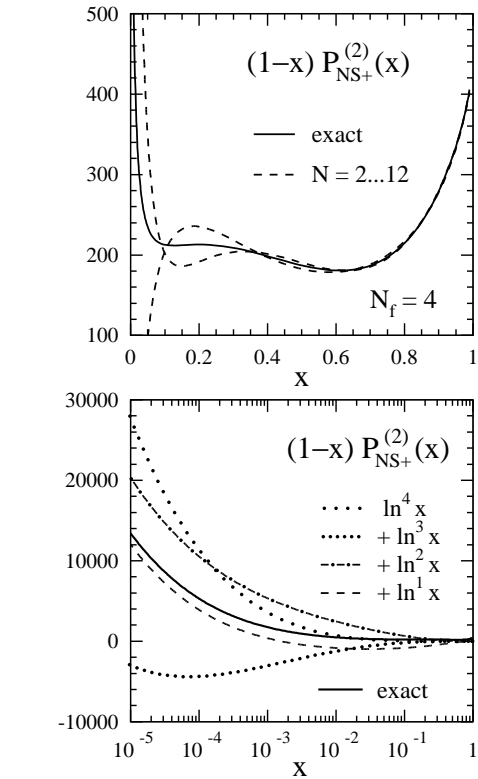}
\caption{One of the three-loop (NNLO) non-singlet splitting functions from Moch, Vermaseren and 
Vogt.\protect\cite{Moch:2004pa,Vogt:2004mw}. See text for details of the various curves.}
\label{fig:pns2nf4}
\end{figure}
One of the three
-loop (NNLO) non-singlet splitting functions from MVV is shown in Fig.~\ref{fig:pns2nf4}. 
In the upper figure, the exact function (solid line) is compared with 
previous approximate forms\cite{vanNeerven:2000wp} fitted to a finite number of exact moments and known $x\to 0,\; 1$ leading behaviour.  In the lower figure, the small-$x$ leading-logarithm, next-to-leading-logarithm, etc. approximations are compared with the exact result.

\begin{figure}%1
\epsfxsize190pt
\figurebox{}{}{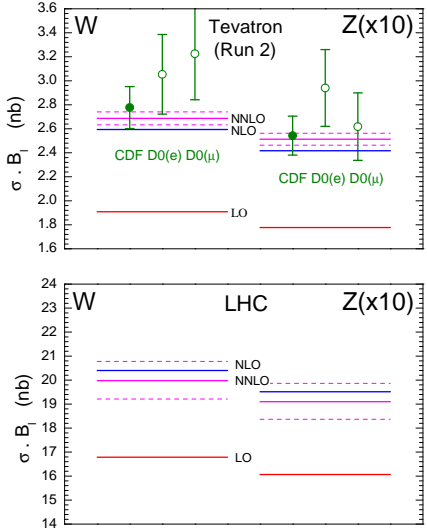}
\vspace*{-1.3cm}
\caption{Predictions for the $W$ and $Z$ total cross sections at the Tevatron and LHC, 
from MRST.\protect\cite{Martin:2002aw,Martin:2003sk}}
\label{fig:wznnlo2004b}
\end{figure}
Using the results of MVV, consistent NNLO calculations of hard-scattering cross sections at hadron colliders can now be performed. One of the most important calculations is the prediction for the total $W$ or $Z$ production cross sections.
With sufficient theoretical precision, these could be used as a {\it luminosity monitor} at the Tevatron or LHC.
A recent NNLO calculation from Martin {\it et al.} (MRST),\cite{Martin:2002aw,Martin:2003sk} which uses the full MVV three-loop splitting  functions, is shown in Fig.~\ref{fig:wznnlo2004b}, together with recent CDF and D0 Tevatron cross-section measurements.  Taking all sources of uncertainties in the theoretical analysis into account, MRST quote a $\pm 4\%$
uncertainty in the prediction,\cite{Martin:2003sk} shown as the horizontal dashed lines bracketing the NNLO cross sections
in Fig.~\ref{fig:wznnlo2004b}. Evidently the perturbation series is well under control, and there is also good agreement between theory
and experiment.

\begin{figure}%1
\epsfxsize190pt
\figurebox{}{}{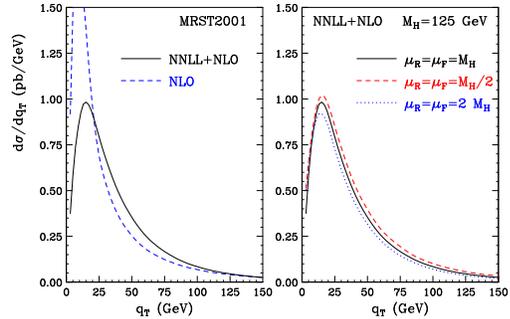}
%\vspace*{-0.5cm}
\caption{Resummed pQCD prediction for the Higgs transverse momentum distribution at the LHC,
from Bozzi {\it et al.}\protect\cite{Bozzi:2003jy}}
\label{fig:nnllpaper}
\end{figure}
\section{Resummation}\label{sec:resum}

Effort continues to refine the predictions for `Sudakov' processes, of which the most familiar examples are the 
thrust distribution near $T=1$ in $e^+e^-$ collisions, and the $W$, $Z$ or Higgs transverse momentum distribution at small
$q_T$ in hadron-hadron collisions. In the latter case, a precise knowledge of the $q_T$ distribution is necessary in order to be able to accurately measure the properties (mass, width, etc.) of the heavy boson. At small $q_T$ the perturbation series becomes dominated by terms of the form
$\sum_{n,m\leq 2n-1} \alpha_S^n \log^m(M^2/q_T^2)$ and resummation is necessary. To obtain a prediction valid over the {\it full}
$q_T$  range, the resummed cross section must be matched onto the exact fixed-order calculation at large $q_T$. This is not trivial, since the resummed calculation ignores power-law contributions of order $(q_T^2/M^2)^N$ that contribute
at large $q_T$. Matching is achieved by first subtracting from the full fixed-order calculation those logarithmic
contributions that form part of the resummed calculation, and then adding the two types of contribution. 
Figure~\ref{fig:nnllpaper} shows a recent calculation of the Higgs $q_T$ distribution at the LHC from Bozzi {\it et al.} which combines the fixed-order (NLO) and resummed (NNLL) contributions. The effect of resummation is clearly seen at 
small $q_T$ in the left-hand figure. The right-hand figure shows that the scale independence of the fixed-order result is maintained in the full calculation over the whole $q_T$ range.

%
%\begin{figure}%1
%\epsfxsize190pt
%\figurebox{}{}{klhc.ps}
%%\vspace*{-0.5cm}
%\caption{Summary of $\asmz$ measurements, from S. Frixione et al.\protect\cite{Catani:2003zt}.}
%\label{fig:klhc}
%\end{figure}

\section{The CSW technique for QCD amplitudes}

Although the numerical calculation of tree-level QCD scattering amplitudes {\it can} be automated, the method is very much `brute force' and the complexity associated with multiparton amplitudes soon saturates the computer capability.
And as already mentioned, equivalent automation for loop amplitudes seems some way off. Analytic expressions are of course only feasible for few-particle amplitudes, and lengthy analytic expressions are in any case of little practical
use for numerical computations: see, for example, the expressions for the three-loop splitting functions in 
Refs.~\cite{Moch:2004pa,Vogt:2004mw}.

\begin{figure}%1
\epsfxsize190pt
\figurebox{}{}{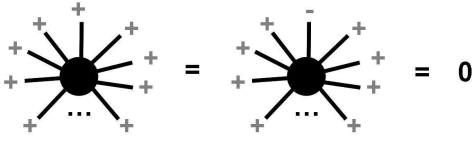}
%\vspace*{-0.5cm}
\caption{Vanishing $n$-gluon helicity amplitudes.}
\label{fig:MHV1}
\end{figure}
\begin{figure}%1
\epsfxsize150pt
\figurebox{}{}{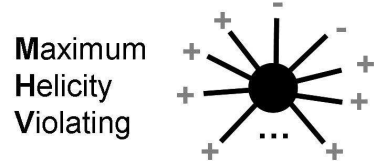}
%\vspace*{-0.5cm}
\caption{The MHV amplitude.}
\label{fig:MHV2}
\end{figure}

In the past year, a new technique based on an idea by Witten\cite{Witten:2003nn} and subsequently developed by Cachazo, Svrcek and Witten\cite{Cachazo:2004kj} (CSW)
may be the long-awaited breakthrough in making QCD multiparticle tree-level and loop amplitudes tractable.\footnote{The Parallel Session contribution of Zvi Bern gives a very clear and comprehensive introduction to the CSW method.}
The discussion of the CSW method starts with an observation by Parke and Taylor\cite{Parke:1986gb} in 1986 concerning the helicity dependence of $n$-gluon QCD tree-level scattering amplitudes. If all the external gluon helicities are the same, or if only one has a different
sign, then the amplitude vanishes, see Fig.~\ref{fig:MHV1}. Although other amplitudes are non-zero in general, the Maximum Helicity Violating (MHV) amplitude, in which two and only two external gluons have the opposite helicity 
(see Fig.~\ref{fig:MHV2}) appears to have a special status in that it can be written in an extremely compact form in
terms of spinor products:
\begin{equation} 
{\cal A}_{\rm MHV} = i\; g_S^{n-2}\; {  \langle r,s \rangle^4 \over  \prod_{j=1}^n \langle j, j+1\rangle}
\label{eq:MHV}
\end{equation}
where colour factors have been suppressed, and the external momenta are labelled $i=1, ...,n$ with $r$ and $s$ the two lines with opposite helicity. No other non-zero helicity amplitudes can be written in such a simple form. The $\langle\; \rangle$ quantities in (\ref{eq:MHV}) are spinor products defined by
\begin{equation} 
\langle i,j\rangle = u_-(p_i) \bar u_+(p_j)\; , \ \ \vert \langle i,j \rangle \vert 
= \sqrt{2p_i \cdot p_j}
\label{eq:spinors}
\end{equation}
The spinor product $ \langle i,j\rangle $ is a complex function of the two four-momenta $p_i$ and $p_j$, the exact form depending on the choice of spinor representation.
For example, a convenient choice for programming purposes is\cite{Kleiss:1985yh}
\begin{eqnarray} 
\langle i,j\rangle &=& (p_j^y - i p^z_j ) \sqrt{\frac{p_i^0 - p_i^x}{p_j^0 - p_j^x}} \nonumber \\
&& - (p_i^y - i p^z_i ) \sqrt{\frac{p_j^0 - p_j^x}{p_i^0 - p_i^x}}
\end{eqnarray}
for momentum 4-vectors $p^\mu = (p^0,p^x,p^y,p^z)$.

\begin{figure}%1
\epsfxsize190pt
\figurebox{}{}{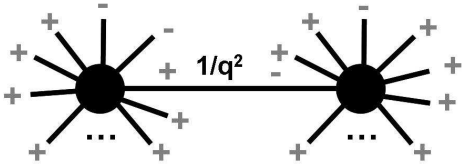}
%\vspace*{-0.5cm}
\caption{Construction of a non-MHV amplitude from two MHV amplitudes and a scalar propagator.}
\label{fig:MHV3}
\end{figure}

In the CSW method, the MHV amplitudes are elevated to {\it effective vertices} in a new scalar graph approach to tree-level
amplitude computation. Other helicity amplitudes are obtained by stitching together MHV amplitudes with scalar propagators, taking account of all possible permutations of helicities. As an example, Fig.~\ref{fig:MHV3} shows how the $(---+++...)$ amplitude is obtained by combining two $(--+++...)$ MHV amplitudes with a $1/q^2$ scalar propagator. The result is a compact expression written in terms of spinor products (generalised to allow for off-shell momenta) much simpler than the corresponding output obtained using the automated programmes. It is of course trivial to check that the CSW-method amplitudes agree with those obtained using standard techniques.
It is also possible to develop a recursive technique that allows MHV vertices to be stitched together in a systematic way in order to build up more complicated amplitudes.\cite{Bena:2004ry}

Two immediate questions are: (i) why does the method `work', and (ii) can it be extended to more general QCD amplitudes, including those with loops? The answer to (i) lies in the fact that
the perturbative expansion of ${\cal N}=4$ Supersymmetric Yang-Mills (SYM) gauge theory  is equivalent to the instanton expansion of a certain type of string theory in supersymmetric 
twistor space.\cite{Witten:2003nn}. Tree-level amplitudes with $n+1$ negative helicity gluons are related to $D-$instantons of degree $n$. The MHV vertices are localised on a line in twistor space,
and the lines are connected by off-shell propagators. Because of this topological structure, amplitudes in unbroken gauge theories
are therefore much simpler than their Feynman diagram expansion would suggest. Since
${\cal N}=4$ SYM gauge theory shares the same gauge boson amplitudes as pure Yang-Mills theory, the results for gluonic amplitudes in QCD follow immediately.

Since the original CSW paper, there has been a steadily increasing amount of theoretical 
activity.\cite{Bena:2004ry,Georgiou:2004wu,Kosower:2004yz,Georgiou:2004by,Khoze:2004ba,Cachazo:2004zb,Brandhuber:2004yw}
The method has been extended to amplitudes with fermions\cite{Georgiou:2004wu}, and the important question of whether there is any equivalent simplification for QCD amplitudes containing loops has begun to be addressed\cite{Cachazo:2004zb,Brandhuber:2004yw}. While the MHV method appears to work at the one-loop level in ${\cal N}=4$ SYM, the extension to loop amplitudes in {\it non-supersymmetric} theories auch as QCD appears to be more problematic. Nevertheless given the current level of theoretical activity, there is likely to be much progress in the near future.

\section{Diffractive Higgs production and related processes}

The discovery of the Higgs boson is obviously one of the main goals of the LHC. For the `standard'
search scenario, $pp\to HX$, via $gg\to H$, $q\bar q \to WH$, $gg\to t\bar t H$ etc., the theory rests on the solid 
foundation of the QCD factorization theorem, and the total rates and kinematic distributions can be accurately
predicted for a given $M_H$. There is, however, no single optimum detection process but rather a range of
possibilities depending on $M_H$, none of which is compelling on its own. It is important, therefore, to explore
other `non-standard' search scenarios.
One that has received much attention recently (for a review of recent work and a list of references, see 
Ref.~\cite{Martin:2004pi})
is the exclusive process $p p \to p \oplus H \oplus p$, where the $\oplus$ signals the presence of a rapidity gap. Provided that appropriate forward proton taggers can be installed, it may be possible
to measure the $M_H$ to high precision via the missing mass associated with the forward protons. This precision would then enhance the signal over the continuum background, for example in the $H\to b \bar b$ channel.  Furthermore, the signal to
background is also enhanced by a $J_z = 0$, P-even selection rule that suppresses the $gg\to b \bar b$ 
background.\cite{Khoze:2000jm,DeRoeck:2002hk} Although the Standard Model Higgs boson is the obvious candidate for a new particle search in this channel, other exotic production processes are possible. Examples are SUSY Higgs bosons, gluino bound states, gravitons etc., indeed anything that couples strongly to gluons and survives the selection rule.\cite{Khoze:2001xm}

\begin{figure}%1
\epsfxsize150pt
\figurebox{}{}{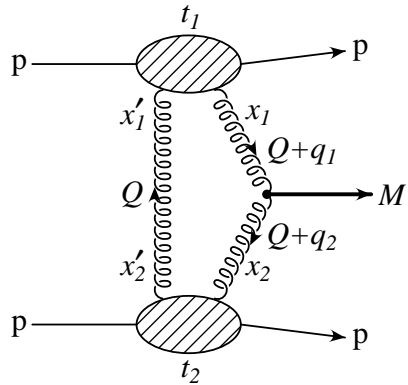}
%\vspace*{-0.5cm}
\caption{Schematic diagram for exclusive Higgs production at the LHC, $p p \to p \oplus H \oplus p$.}
\label{fig:vak1}
\end{figure}

However the calculation of processes such as $p p \to p \oplus H \oplus p$ presents a real challenge for the theory, since it involves both perturbative and non-perturbative aspects. In particular, one needs techniques for calculating not only the production amplitude shown in Fig.~\ref{fig:vak1},
but also the rapidity gap `survival probability', i.e. the probability that strong interactions between the upper and lower
proton systems in Fig.~\ref{fig:vak1} do not populate the rapidity interval between them with additional soft hadrons. Preliminary indications are that signal to background ratios greater than one can indeed be achieved,\cite{Khoze:2001xm} and the possibility of measuring similar processes at the Tevatron (for example, $p \bar p \to p \oplus \chi_c \oplus \bar p$) offers a means of checking and calibrating the Higgs calculation. On the experimental side, the missing mass resolution is crucial with 
$\Delta M_{\rm miss} \sim 1\;$GeV being the goal, see for example the recent study of Ref.~\cite{Boonekamp:2004nu}.

\vspace*{-0.25cm}

\section*{Acknowledgements}
I would like to thank all the conveners and speakers in the relevant parallel sessions, and also colleagues
at the IPPP, for their help in preparing this review. Siggi Bethke kindly provided his latest $\alpha_S$ compilation. I am grateful to the Royal Society for financial support
in the form of a travel grant. Finally, the conference organisers are to be congratulated for making ICHEP04 such
an enjoyable and stimulating meeting.
%\vspace*{-0.25cm} 

\end{document}